\documentclass[fleqn,10pt]{wlscirep}
\usepackage[percent]{overpic}

\title{Towards `smart lasers': self-optimisation of an ultrafast pulse source using a genetic algorithm}

\author[1,*]{R. I. Woodward}
\author[1]{E. J. R. Kelleher}
\affil[1]{Femtosecond Optics Group, Department of Physics, Imperial College London, London, UK}

\affil[*]{r.woodward12@imperial.ac.uk}

\begin{abstract}
Short-pulse fibre lasers are a complex dynamical system possessing a broad space of operating states that can be accessed through control of cavity parameters. Determination of target regimes is a multi-parameter global optimisation problem.
Here, we report the implementation of a genetic algorithm to intelligently locate optimum parameters for stable single-pulse mode-locking in a Figure-8 fibre laser, and fully automate the system turn-on procedure.
Stable ultrashort pulses are repeatably achieved by employing a compound fitness function that monitors both temporal and spectral output properties of the laser.
Our method of encoding photonics expertise into an algorithm and applying machine-learning principles paves the way to self-optimising `smart' optical technologies.
\end{abstract}

\begin{document}

\flushbottom
\maketitle

\thispagestyle{empty}

As the importance of ultrafast laser sources for industrial, scientific and medical applications continues to grow, so too do the demands for increasingly versatile and reliable systems, driving research in this field towards highly engineered solutions.
Passively mode-locked fibre lasers are particularly attractive as sources of femtosecond and picosecond pulses due to their compact footprint, robust construction, excellent heat dissipation capability allowing power scaling, and superior beam quality.

Passive mode-locking is achieved using an intracavity saturable absorber.
While significant progress has been made in the development of new materials as \emph{real} ultrafast saturable absorbers~\cite{Martinez2013a,Woodward2015_as_2d,Sobon2015}, their response time and operating wavelength range remains inferior to \emph{artificial} saturable absorber technologies---schemes that exploit third-order nonlinear effects in glass fibre (reacting on femtosecond timescales, and largely independent of wavelength) to induce an intensity-dependent transmission. 
Examples of these include nonlinear polarisation evolution (NPE) and nonlinear loop mirror schemes~\cite{Duling1991,Hofer1991}.
Artificial saturable absorber technologies are also often simpler and more cost effective to implement, without the need for complex materials processing and growth.
Additionally, in contrast to the stationary nonlinear response of a real saturable absorber, the nonlinear transfer function of an artificial saturable absorber can be tuned by adjusting various system parameters (e.g. polarisation and optical power). 
Adjustment of the effective nonlinear transfer function supports traversal of a broad range of lasing states, including continuous-wave, Q-switched, mode-locked, and a variety of intermediate or unstable noisy pulsation regimes.
While this feature has enabled the exploration of distinct regimes and revealed rich nonlinear physics with analogues in other dynamical physical systems~\cite{Lecaplain2012,Churkin2015,Woodward_2016_pre,Runge2015,Haefner2016}, determination and control of parameters to maintain a desired mode of operation is a multi-parameter global optimisation problem. 
The optimisation is further complicated by the influence of external perturbations (e.g. thermal and mechanical effects) to the system, leading to a non-stationary solution that varies unpredictably as a function of time and temperature etc., unless the operating environment is mechanically and thermally stabilised.
This issue is a major limitation preventing widespread application of mode-locked fibre lasers that employ artificial saturable absorbers, while the ability to engineer self-starting systems that operate reliably, without specialist user intervention, remains an open problem. 

A promising solution is the application of automated electronic control systems to tune the laser parameters. 
At the simplest level, this can involve linearly sweeping electronically controlled parameters while monitoring the output, waiting for a desired regime to be found~\cite{Hellwig2010,Olivier2015}, followed by a feedback loop to maintain this state in the presence of disturbances~\cite{Shen2012,Radnatarov2013,Brunton2013}.
This procedure is slow, however, and quickly becomes intractable when multiple laser variables are included that increase the dimensionality of the parameter space.
Additionally, the feedback system performs only local optimisation and may prevent a superior operating regime from being identified once the system locates a \emph{local} maxima.
A universal solution, applicable to different laser designs, should find the \emph{globally} optimum operating regime without any prior knowledge of the system.
To achieve this, global multi-parameter optimisation can be efficiently implemented using machine learning principles.
Genetic algorithms (GAs) are ideal for this task, applying concepts of natural selection from evolutionary biology to intelligently search for optimum parameters~\cite{Melanie1996}.
This technique has previously been applied in a number of optical contexts, such as pulse shaping~\cite{Baumert1997}, optimisation of supercontinuum generation~\cite{Arteaga-Sierra2014}, and the design of specialist optical fibres and amplifiers~\cite{Prudenzano2007,Kerrinckx2004}.
The application of GAs for extremum seeking in laser mode-locking was recently proposed theoretically~\cite{Fu2013,Brunton2014}, and a basic implementation utilizing a singular fitness function to locate self-starting regimes in an NPE-mode-locked fibre laser was demonstrated~\cite{Andral2015,Andral2016}.
It was observed, however, that coherent single-pulse mode-locking was not repeatable, and optimisation of the fitness function led to a tendency towards the emission of noise-like pulses, highlighting the need to implement a more sophisticated fitness function~\cite{Woodward_cleo16_ga}.

Here, we experimentally demonstrate the first photonic application of a GA based on a \emph{compound} fitness function to achieve optimised, reliable self-starting operation of an all-fibre ultrafast pulse source, paving the way towards fully automated `smart lasers'. 
We show that the self-optimising scheme can rapidly explore a large multi-parameter space, locating and maintaining a global optimum in the presence of external disturbances.

\section*{Self-Optimising Laser Design}
\subsection*{Operating States in Fibre Lasers}
We consider a Figure-8 (F8) laser design: one of the earliest reported passively mode-locked laser schemes, employing a nonlinear amplifying loop mirror (NALM) as an artificial saturable absorber~\cite{Duling1991}, and currently receiving renewed interest as a flexible, low-cost all-fibre ultrafast source~\cite{Nicholson2007a,Pottiez2010,Zhao2013b,Runge2014a,Szczepanek2015,Hao2016,Xu2014a}.
F8 lasers consist of a unidirectional and bidirectional ring. 
The bidirectional ring forms a loop mirror that is imbalanced (either actively or passively) to induce a differential phase, and consequently a power-dependent reflectivity that mimics the action of a saturable absorber, promoting pulse generation in the main laser cavity. 
Polarisation control (PC) that acts as a phase bias is often included in the loop mirror to adjust the nonlinear transfer function of the NALM, and consequently the effective saturable absorber behaviour.
Additional to polarisation control, the variable gain from an amplifier in the loop mirror, and influences from external disturbances (e.g. thermal and mechanical stresses) can affect the differential phase of the counter propagating waves and modify the nonlinear transfer function.    
Thus, to achieve stable self-starting operation in a target regime (e.g. single-pulse mode-locking), the phase bias must be carefully set and actively controlled.

\begin{figure}[ht]
	\centering
	\includegraphics{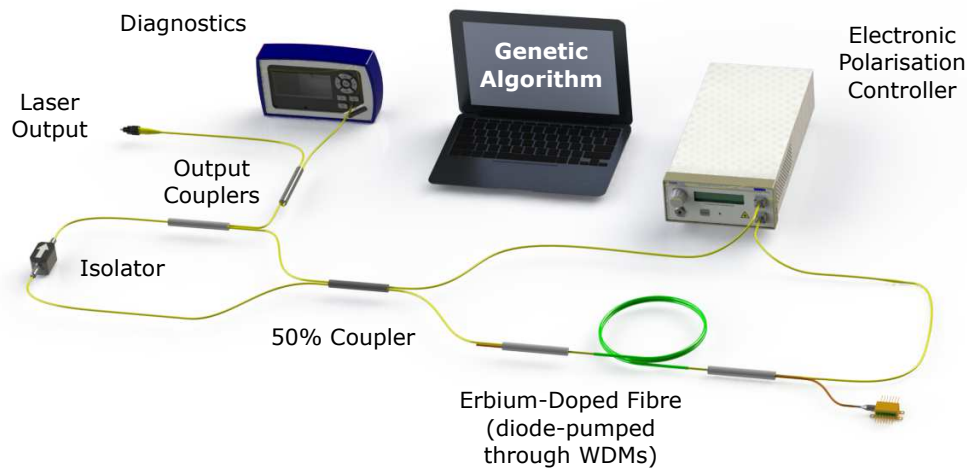}
	\caption{Cavity schematic of self-optimising mode-locked laser.}
	\label{fig:cavity}
\end{figure}

\begin{figure}[ht]
	\centering
	\sffamily
	\begin{overpic}{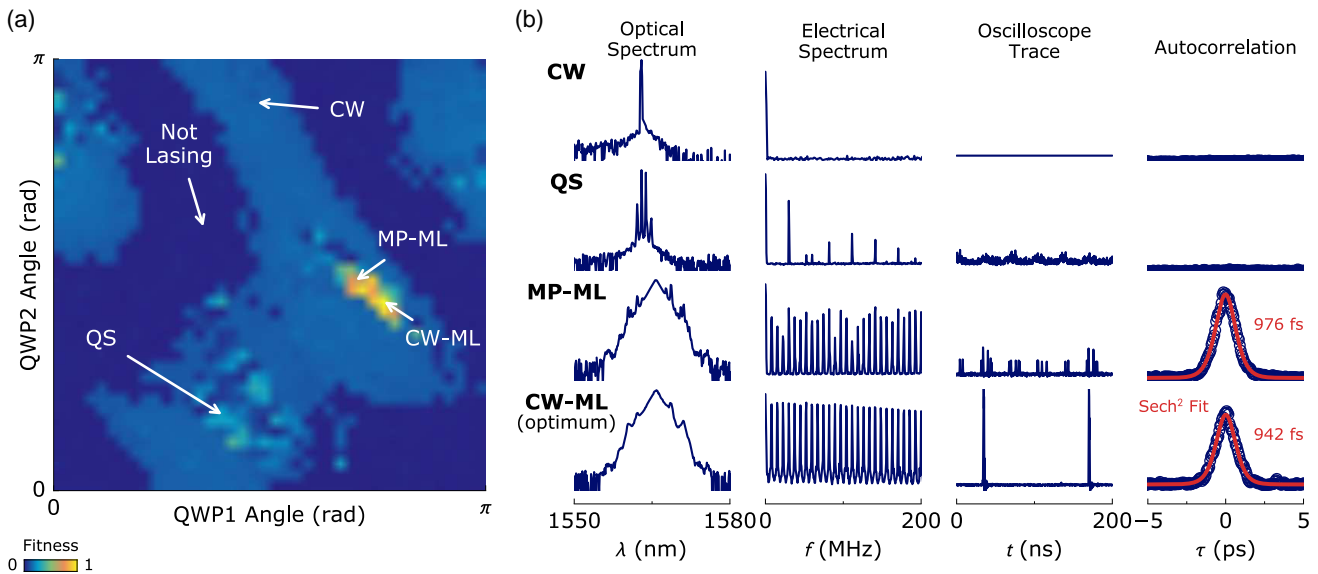}
		\put(0,42){{\small (a)}}
		\put(39, 42){{\small (b)}}
	\end{overpic}
	\normalfont
	\caption{(a) Map of laser output fitness score (where 0 indicates no lasing and 1 indicates optimally stable single-pulse mode-locking) as a function of waveplate position. (b) Typical output properties of salient operating states, where the y-axis is intensity, with a log scale for the spectra and linear scale for temporal diagnostics. $\lambda$ - wavelength, $f$ - frequency, $t$ - time, $\tau$ - delay.}
	\label{fig:heatmap}
\end{figure}

The F8 laser setup is shown in Fig.~\ref{fig:cavity} (see Methods section for details). 
The nonlinear amplifying loop mirror includes an electronic polarisation controller with four quarter waveplates (QWPs), providing full-wave control and the ability to traverse the entire surface of the Poincar\'{e} sphere. 
For a fixed pump power, the nonlinear transmission curve of the NALM is governed by the angle of the four waveplates, adjustment of which sweeps the laser operating regime through a wide range of states.
To illustrate this variation, we represent a two-dimensional slice of the four-dimensional polarisation space: the pump power is held constant while two quarter waveplates are successively swept through all possible angles in 4.5~degree (0.025$\pi$ rad) steps (Fig.~\ref{fig:heatmap}).
At each angle the output properties of the laser are evaluated and assigned a `fitness score' quantifying the laser performance, discussed in detail later.

Localised regions of highest fitness indicate stable single-pulse continuous-wave mode-locking (CW-ML) at a repetition rate that matches the fundamental cavity frequency of 7.4~MHz. 
The lowest scores are assigned to non-lasing states and CW emission. 
Intermediate fitness values represent a wide variety of pulsating regimes, including Q-switching (QS) and unstable multiple-pulse or partial mode-locking (MP-ML).
The characteristic output properties, evaluated in the optical and electrical spectral and temporal domains, that indicate operation in these regimes are summarised in the right-hand panels of Fig.~\ref{fig:heatmap}.
Evident from Fig.~\ref{fig:heatmap} is the sparseness of stable states, highlighting the need for an extremum seeking approach to quickly and efficiently find optimum regimes from any unknown initial state.

\subsection*{Genetic Algorithm Development}
GAs are well-suited to the task of finding optimum solutions to a multi-parameter problem, where the quality of a solution is measured by a fitness function that is dependent on the value of selected system variables.
In the nomenclature of GAs, each possible solution is known as an `individual' and comprises a set of values for each parameter. 
Individual parameters are referred to as `genes'~\cite{Melanie1996}.

\begin{figure}[ht]
	\centering
	\includegraphics{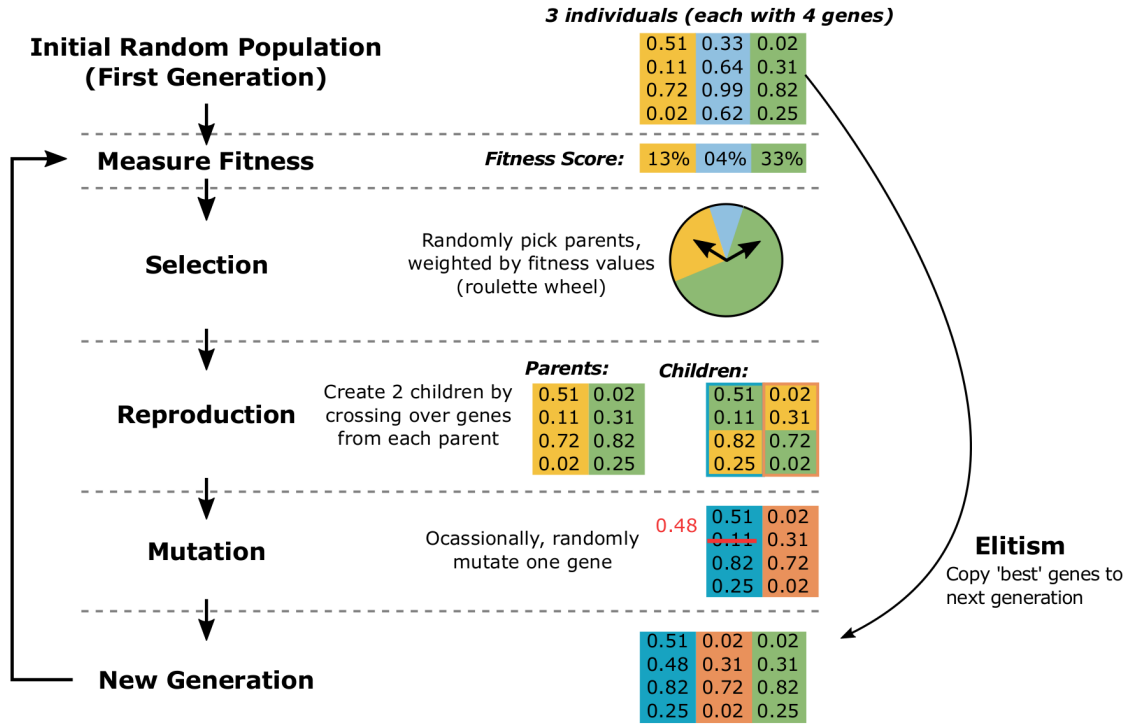}
	\caption{Illustration of the genetic algorithm concept, showing an example iteration of the algorithm with a population of three individuals, each consisting of four genes.}
	\label{fig:ga_explained}
\end{figure}

Fig.~\ref{fig:ga_explained} illustrates the core GA concept, which we now briefly explain.
The process begins with a collection of individuals, each comprising a set of randomly assigned genes.  
This group (or `population') becomes the first generation and represent the evolutionary epoch.
The system output is measured for each individual in the generation---in our case by electronically setting cavity parameters based on the individual's genes.
This output is evaluated by a fitness function (also known as a merit or objective function) and assigned a score.

The GA then creates the next generation by breeding individuals from the preceding generation, with the probability that an individual is selected to be a `parent' based on their score (`roulette wheel' selection~\cite{Melanie1996}).
Two children are produced from two parents by randomly distributing the parent's genes between the children.
A mutation probability is also specified, which can randomly alter the children's genes.
This process repeats until a steady state is reached, with a high probability of retaining good genes in the population, but permitting diversity to locate global maxima through breeding and mutation, while breeding out low quality genes which result in a low fitness score since they have low selection probability for breeding.
The principle of `elitism' is also employed, cloning the best individuals to the next generation to ensure their high-quality genes are preserved, increasing the speed of convergence~\cite{Melanie1996}.

We find that a generation size of 30 individuals, each consisting of five genes (four waveplate angles and pump current), provides a suitable balance between genetic diversity and speed of convergence. 
The inclusion of pump current (that is proportional to pump power above the pump laser threshold) as a gene permits the full turn-on cycle and self-starting behaviour to be intelligently automated.
The range of allowed pump currents is bounded: the upper limit prevents damage to integrated optical components and the pump laser itself (a 965~nm fibre-pigtailed diode laser); while the lower bound ensures the pump diode is above threshold.   
A damped mutation is also applied, where an initial value of 25\% is adaptively reduced as the algorithm progresses to assist convergence. 
We also apply elitism by always cloning the best four individuals from each generation to the next.
Our choice of GA parameters was empirically determined to give repeatable and reliable results; further optimisation to maximise the convergence rate is possible, but beyond the scope of this investigation~\cite{Melanie1996}. 

\subsection*{Choice of Fitness Function}
A critical factor to the success of a self-optimising laser implementation is the fitness function, which must return a higher value when the laser is operating closer to the target regime. 
In previous work, using local search algorithms, a variety of approaches have been considered to evaluate laser performance.
Shen et al.~\cite{Shen2012} counted the optical pulses arriving on a photodiode to determine if the desired mode-locked repetition rate had been achieved.
Olivier et al.~\cite{Olivier2015} monitored the output polarisation state, detecting an abrupt change as a transition into mode-locking as cavity waveplate angles were swept.
Radnatarov et al.~\cite{Radnatarov2013} quantified high quality mode-locking by maximizing the value of the fundamental RF beat note in the electrical spectrum.
To identify high-peak-power short-pulse regimes, nonlinear optical processes, including two-photon absorption (TPA)~\cite{Hellwig2010} and second-harmonic generation (SHG)~\cite{Andral2015}, have been employed. 
It has been found, however, that in isolation these singular fitness functions are unable to fully determine the laser output state and thus reliably enable self-optimisation, in particular when identifying single-pulse CW mode-locking.
For instance, Q-switched mode-locked regimes can also produce high-peak-power pules leading to a high SHG signal~\cite{Andral2015}. 

In practice, a specialist user assesses laser performance through a variety of diagnostics to confirm specific modes of operation, combing information from both the temporal and spectral domains.
Therefore, we propose that a compound fitness function, including optical and electrical spectral, and temporal measurements, is required for reliable self-optimisation~\cite{Woodward_cleo16_ga}.
While this approach could be applied for a wide range of target output properties, here, we focus on ultrafast single-pulse CW mode-locking.
Such pulse sources are an enabling tool in a wide range of applications including manufacturing, research and medical imaging, where pulse to pulse stability, and reliability of the system are of critical importance.

A stable train of mode-locked pulses incident on a photodiode induces a periodic electrical signal.
Visualised on an oscilloscope, single-pulse mode-locking is characterised by pulses regularly spaced by the round-trip time of the laser cavity. 
Represented on an electrical spectrum analyser, the Fourier transform of the output exhibits bands appearing at the fundamental cavity frequency ($f_\mathrm{rep}$), and harmonics thereafter ($2f_\mathrm{rep}$, $3f_\mathrm{rep}$, etc).
The magnitudes of these bands in the power spectrum monotonically decrease with the frequency response of the detection, while the signal-to-noise ratio (SNR) is a widely used metric for assessing the stability and quality of mode-locking~\cite{Linde1986}.
Instabilities including Q-switched mode-locking and multiple pulsing reduce the peak to pedestal contrast or introduce lower frequency envelope modulations of the spectral power.
In terms of their optical spectrum, Fourier transform limited pulses have a duration that is inversely proportional to their spectral bandwidth, thus ultrafast lasers are characterised by a broad spectrum on an optical spectrum analyser. 
In combination these three diagnostics provide a comprehensive picture of stable trains of ultrashort pulses in both the time and frequency domain. 

We use this information to formulate a compound fitness function, as follows: 
\begin{itemize}
\item An oscilloscope score is assigned based on maximising the peak of a single pulse observed on a time-scale that corresponds to a single round-trip, where multiple-pulsing regimes are readily identified by numerous lower-intensity pulses present in this window.
\item An electrical spectrum score is assigned based on maximising the magnitude of the fundamental frequency, penalising amplitude fluctuations in the harmonic frequencies up to 500~MHz.
\item An optical spectrum score is based on maximizing the laser 3 dB bandwidth.
\end{itemize}

The total fitness function of an individual is given by the sum of these three components, with equal weighting. 
Further optimisation could be possible by unequal weighting, which is a topic of ongoing work.
We note that ultrafast temporal characterisation using an autocorrelator would provide additional information to enhance this score, however, the polarisation-sensitivity of the underlying nonlinear prohibits its application to lasers with a non-stationary output polarisation.

\begin{figure}[ht]
	\centering
	\includegraphics{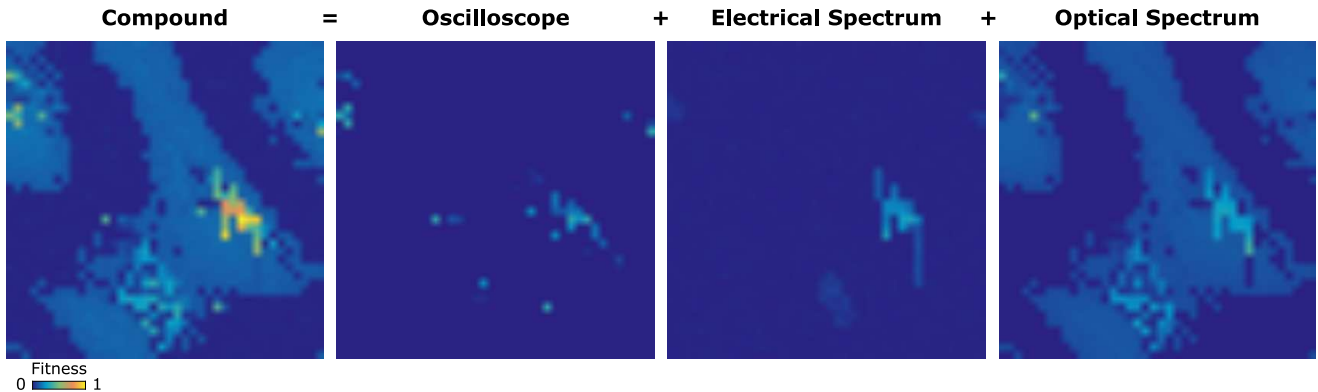}
	\caption{Map of laser output fitness score, decomposed into components forming the compound fitness function (x and y axes are QWP1 and QWP2 angle, respectively, swept through 180 degrees). The data was recorded under the same conditions as Fig.~\ref{fig:heatmap}, but without resetting the waveplates for each measurement (i.e. including hysteresis effects).}
	\label{fig:heatmap_exploded}
\end{figure}

The component contributions to the total fitness are shown in Fig.~\ref{fig:heatmap_exploded}.
It is clear that the composite fitness function provides highest contrast between stable and unstable pulsating regimes. 
In contrast the singular fitness functions fail to distinguish the subtleties between mode-locking regimes. 
The optical spectrum clearly distinguishes between lasing and non-lasing regions (i.e. when the NALM phase bias results in a very high cavity loss, or the pump power is too low), which is particularly important in the early GA evolution. 
The optical spectral width is also found to be a reliable metric for distinguishing between Q-switched and mode-locked regimes, although it fails to differentiate between multiple pulsing, Q-switched mode-locking and stable CW mode-locking, since they all yield broad spectra (noting that the optical spectrum analyser inherently averages over many consecutive pulses).
Measurement of spectral harmonics in the electrical domain clearly highlights mode-locked regimes, although this alone has already been proven insufficient to preferentially locate stable single-pulse mode-locking~\cite{Andral2015}.
The oscilloscope is therefore important for identifying and promoting fundamental, single-pulse mode-locking over multiple pulse mode-locking, but in isolation is a poor metric in the early stages of the evolution.

Adopting a thermodynamic picture, mode-locking can be viewed as a first-order phase transition that occurs at a critical `temperature' or threshold power, where an initially long-lived, metastable CW regime finally transitions into the stationary mode-locked state~\cite{Gordon2006}. 
This dynamic leads to a hysteresis behaviour exhibited by the system: mode-locking persists over a wider range of parameter space when already in the stable state, but not all mode-locked states yield self-starting behaviour.
We explore the impact of this phenomenon by performing a 2D parameter sweep of the intra-cavity polarisation with (Fig.~\ref{fig:heatmap}) and without (Fig.~\ref{fig:heatmap_exploded}) resetting the laser state to a CW regime between each parameter adjustment.
Firstly, we note that the fitness contours are highly repeatable for successive sweeps performed at near-constant ambient temperature.
Lines of higher fitness can be seen extending vertically from the mode-locking region in Fig.~\ref{fig:heatmap_exploded} (the vertical lines are explained by the fact that the y-axis waveplate is stepped from 0 to $\pi$, followed by an increment of the x-axis waveplate; this process repeats across the two-dimensional parameter space), which are absent when we reset to a CW state between parameter cycles (Fig.~\ref{fig:heatmap}).
Although the hysteresis is evident, the effect is small and the macroscopic pattern of stable operation remains largely unchanged.
In addition, while a high scoring regime may be identified that occurs due to hysteresis in the system, and as such does not represent a reliably self-starting state, the random dependence of the GA on the previous system state will ensure that the algorithm will favour states which are always repeatable in the long-term.
Thus, in order to increase the speed of convergence, for the proceeding discussion we do not implement parameter resetting.
Similarly, we note that unstable non-stationary pulse trains that instantaneously yield a high-score, but are rapidly transitioning between regimes, will also ultimately be bred out in favour of stable regimes with high repeatability.

\section*{Results and Discussion}
\begin{figure}[ht]
	\centering
	\sffamily
	\begin{overpic}{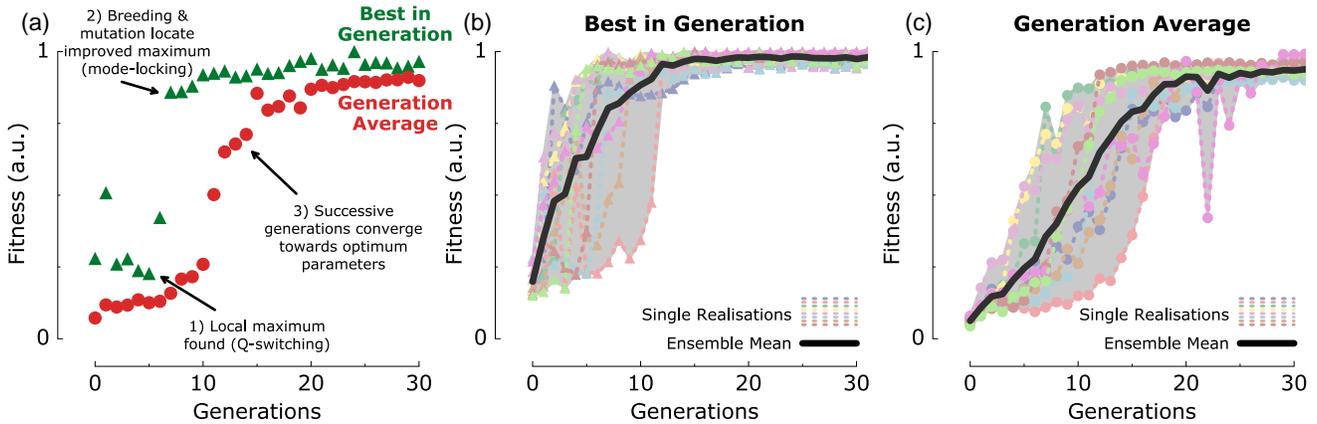}
		\put(1,30){{\small (a)}}
		\put(35, 30){{\small (b)}}
		\put(69,30){{\small (c)}}
	\end{overpic}
	\normalfont
	\caption{Evolution of fitness score for: (a) single realisation, showing convergence of successive generation's average score towards the maximum; (b) maximal and (c) average fitness values for an ensemble of ten realisations.}
	\label{fig:convergence}
\end{figure}

The compound fitness function-based GA is applied to the Figure-8 laser, initialised from a unique, randomised set of polarisation parameters, and a pump current corresponding to a sub-threshold power. 
The evolution is shown in Fig.~\ref{fig:convergence}(a), highlighting the fitness score of the best individual in the population over successive generations, as well as each generation's average.
As expected, the average fitness of the initial random seed population is low, with the majority of individuals corresponding to non-lasing or CW states and fittest individuals resulting in Q-switched operation.
Successive generations maintain the best individuals through elitism, while breeding new individuals with a high probability of inheriting high-quality genes. 
After $\sim$5 generations the output appears to be converging towards a local maxima (with a score of $\sim$0.25) corresponding to Q-switched operation. 
Due to mutation and crossover, however, during the evaluation of parameters in successive generations the GA locates an improved operating regime (represented by the sudden increase in `best in generation' scores in generations 7 and 8).
Through continued breeding and mutation, the optimum score gradually increases leading to the identification of a stable single-pulse mode-locked state.
In subsequent operations, the average score of each generation converges towards the maximum, indicating that `good' genes are bred throughout successive generations.
We emphasise that the complete turn-on cycle and tuning of cavity parameters to optimise stable, ultrashort pulse operation is thus fully automated.

We verify the repeatability of this approach by performing an ensemble of ten realisations, each from an initially `off' state.
While each realisation resulted in a different evolution trajectory, due to random initial conditions and a probabilistic evolution, the final results are consistently similar, as shown in Fig.~\ref{fig:convergence}(b)\&(c) highlighting each generation's best and average fitness.
It is always observed that the laser converges towards a stable fundamentally mode-locked output, effectively optimising itself into the target operating regime.
An average optimisation process requires $\sim$20 generations [Fig.~\ref{fig:convergence}(c)] and completes in typically less than 30~minutes. 
The convergence time is dominated by a delay in remotely interfacing with the electronic polarisation controller, pump diode controller and the diagnostics, in addition to a short settling period allowing the cavity dynamics to stabilise in response to each new individual.

\begin{figure}[ht]
	\centering
	\sffamily
	\begin{overpic}{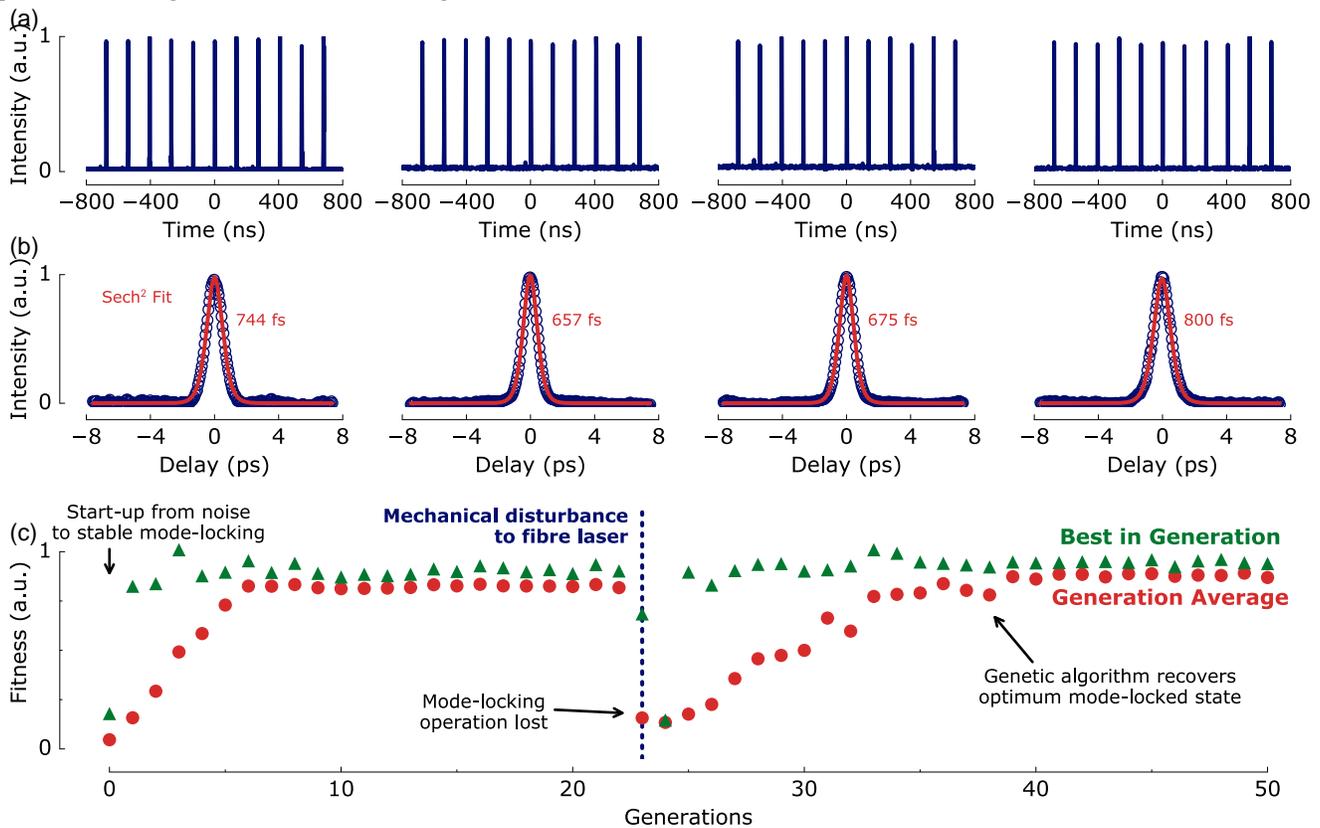}
		\put(0,22){{\small (c)}}
		\put(0,44){{\small (b)}}
		\put(0,61.6){{\small (a)}}
	\end{overpic}
	\normalfont
	\caption{Output properties of laser mode-locked using four consecutive realisation of the genetic algorithm: (a) oscilloscope traces; (b) autocorrelation traces and deconvolved pulse widths. (c) Fitness score evolution showing the genetic algorithm recovers optimum mode-locking after the laser is mechanically perturbed.}
	\label{fig:reliability}
\end{figure}

Suppressing the tendency towards noise-burst and multi-pulsing operation is a common challenge in mode-locked fibre laser design, as highlighted by Andral et al. using an evolutionary algorithm with a single-diagnostic fitness function~\cite{Andral2015,Andral2016}. 
We observe similar unstable multi-pulsing behaviour when the GA is executed with only a singular fitness function as described in Ref.~\cite{Andral2015}.
By using a compound fitness function integrating multiple diagnostics, however, the system repeatability generates stable pulses with sub-picosecond durations (shown by outputs from consecutive realisations in Fig.~\ref{fig:reliability}(a)\&(b), where the autocorrelation is not included in the fitness function but is used to independently verify ultrashort coherent pulse generation).
Variation in the output pulse duration on the order of $\sim$150~fs is noted between realisations. 
This is due to the bandwidth limits of our fitness function diagnostics that prohibit real-time ultrafast characterisation and hence, pulse duration optimisation.
Despite this, we show that the compound fitness function approach is a reliable solution to global optimisation of the laser operating state.
This highlights the improvement from employing multiple diagnostic measurements, albeit at the cost of greater system complexity.
Further progress is therefore possible by integrating femtosecond timescale diagnostics into the fitness function which remains a topic for future work.

Finally, we demonstrate the benefit of online monitoring of the laser output to ensure optimum performance by intentionally disturbing the fibre laser during mode-locked operation. 
Fig.~\ref{fig:reliability}(c) shows a typical evolution from turn-on; in the 23rd generation we mechanically perturb the cavity which alters the fibre birefringence and thus the phase bias of the NALM. 
This disturbance changes the cavity dynamics that the GA has previous `learned' how to optimise and mode-locking is lost.
The subsequent sharp fall in fitness score increases the mutation rate (which was damped when stably operating) permitting the GA to explore a wider parameter space to diversify the population, identifying and ultimately converging towards new optimum parameter settings for stable mode-locking.

\section*{Conclusion}

We have demonstrated a self-optimising ultrafast Figure-8 fibre laser by employing a compound fitness function-based genetic algorithm.
Through exploration of various nonlinear cavity dynamics, which can be accessed by automated control of power and polarisation, we showed that a compound fitness function, assessing both the temporal and spectral output properties of the laser, is required to obtain an accurate `score' for quantifying laser performance. 
This score is maximised by the GA to obtain optimum performance, targeting the generation of stable mode-locked ultrashort pulses.

This approach is ideally suited to mode-locked lasers that employ an artificial saturable absorber, where the nonlinear transfer function can be dynamically controlled electronically. 
More generally, this technique could benefit pulsed lasers including real saturable absorbers, where polarisation and power are still critical parameters affecting stable operation, despite the restricted space of operating regimes due to the fixed nonlinear transfer function of the absorber.
We anticipate our approach could be particularly beneficial in emerging designs of all-normal-dispersion lasers~\cite{Chong2006} and long-cavity lasers~\cite{Woodward_ol_2015_gco,Ivanenko2016}, where the absence of soliton pulse shaping creates a more challenging nonlinear cavity dynamic to achieve stable pulsation, yet has potential for high-energy pulse generation.

To further improve the self-optimising laser design, we expect that the convergence time for the GA could be reduced through careful optimisation of the algorithm parameters (population size, mutation rate etc.) or even by the implementation of automated GA parameter tuning~\cite{Melanie1996}. 
The delay in remote interfacing of diagnostic hardware could also be minimised by electronic integration.
Additionally, while we have focussed on achieving single-pulse mode-locking, a fitness function could be defined to achieve stabilised operation in a wide range of possible laser operating regimes, permitting new studies of the nonlinear cavity dynamics.
Finally, we believe the genetic algorithm approach including a compound fitness function could be widely applicable to photonic device technology, leading to a new generation of intelligent self-optimising systems.

\section*{Methods}
Our Figure-8 laser design [Fig.~\ref{fig:cavity}(a)] includes a passive unidirectional ring and an active bidirectional loop (known as a nonlinear amplifying loop mirror, NALM). 
The passive ring comprises of an isolator and 10\% output coupler, while the NALM contains a 2.3~m length of ytterbium-erbium co-doped fibre, diode-pumped at 965~nm through a wavelength division multiplexer, and an electronic polarisation controller (EPC). 
Our EPC is formed of four stepper motor-controlled fibre-loop quarter waveplates (with 0.18 degree rotation resolution), enabling complete traversal of the Poincar\'{e} sphere by stress-induced birefringence. 
The total cavity length is $\sim$28~m, resulting in a cavity group delay dispersion of $\sim$-0.6~ps$^2$, indicating that soliton pulse shaping is expected.
The cavity output is split to provide a signal to diagnostics for evaluation of the fitness function and to provide the GA-stabilised laser output.


\section*{Acknowledgements}
We thank Roy Taylor, Robert Murray and Timothy Runcorn for stimulating discussions. RIW acknowledges support through an EPSRC Doctoral Prize Fellowship and EJRK is supported by a Royal Academy of Engineering Fellowship.

\section*{Author contributions statement}
Both authors contributed to the experimental design, genetic algorithm development and writing of the manuscript.

\section*{Additional information}

\noindent \textbf{Competing financial interests:} The authors declare no competing financial interests.

\end{document}